*Research Article*

# Stability of Real Parametric Polynomial Discrete Dynamical Systems

## Fermin Franco-Medrano[1,2] and Francisco J. Solis[1]

[1]*Applied Mathematics, CIMAT, 36240 Guanajuato, GTO, Mexico*
[2]*Graduate School of Mathematics, Kyushu University, Fukuoka 819-0395, Japan*

Correspondence should be addressed to Francisco J. Solis; solis@cimat.mx

We extend and improve the existing characterization of the dynamics of general quadratic real polynomial maps with coefficients that depend on a single parameter $\lambda$ and generalize this characterization to cubic real polynomial maps, in a consistent theory that is further generalized to real $m$th degree real polynomial maps. In essence, we give conditions for the stability of the fixed points of any real polynomial map with real fixed points. In order to do this, we have introduced the concept of *canonical polynomial maps* which are topologically conjugate to any polynomial map of the same degree with real fixed points. The stability of the fixed points of canonical polynomial maps has been found to depend solely on a special function termed *Product Position Function* for a given fixed point. The values of this product position determine the stability of the fixed point in question, when it bifurcates and even when chaos arises, as it passes through what we have termed *stability bands*. The exact boundary values of these stability bands are yet to be calculated for regions of type greater than one for polynomials of degree higher than three.

## 1. Introduction

The theory of discrete dynamical systems with iteration functions given by polynomials is an intensive research subject where a wide variety of discrete models have been proposed to describe and to analyze different mechanisms in various areas of science. For example, in Biology and more specifically in Population Dynamics there are many simple models that are used to study the asymptotic behavior of some species that live in isolated generations; see, for instance, [1–7].

Although the dynamics of parametric polynomial discrete systems are very complex their bifurcation diagrams have proved to be a very useful visual tool. A new method for constructing a rich class of bifurcation diagrams for unimodal maps was presented in [8], where the behavior of quadratic maps was analyzed when the dependence of their coefficients was given by continuous functions of a parameter. Conditions on the coefficients of the quadratic maps were given in order to obtain regular reversal maps. Our first goal is to restate the results for more complex systems (cubic) than the quadratic systems analyzed in [8] and to state the results in the frame of a new formulation that would allow for generalization. Our second goal is to generalize the existing results on real quadratic maps for arbitrary real polynomial maps within a framework that allows us to understand the dynamics for a larger set of discrete systems. It is important to remark that our results are analytical and depend only on the parametric derivative of the system evaluated at equilibrium points. There are diverse results based on other approaches such as the linearized stability due to Lyapunov; see, for instance, [9, 10]. In our opinion, our approach is natural for polynomial iteration functions whereas the linearized stability can be used for more complex discrete systems with iteration functions such as piecewise functions. It is also important to notice that there is diverse numerical software specialized in the numerical continuation and bifurcation study of continuous and discrete parameterized dynamical systems, such as Auto [11] and MatCont [12].

Before attempting to obtain general results for polynomial discrete systems, we want to motivate them with those for a nontrivial system. To do this, we propose in Section 2 analyzing the stability of a general cubic discrete dynamical system. Then in Section 3, we use a general framework in order to analyze the stability for real polynomial discrete dynamical systems by using some of the ideas introduced in the previous section. In order to illustrate the obtained results several examples are included in Section 4. Finally, conclusions are given in Section 5.

## 2. Cubic Discrete Dynamical Systems

To motivate our results, we will start with a general cubic discrete system since it is in this case, besides linear and quadratic systems, when explicit calculations can be achieved. Then, for such system, we will define two particular forms, namely, the *Linear Factors Form* and the *Canonical Form*. We will show that these two forms are actually topologically conjugate, which in turn means that the property of chaos is preserved between the maps, which allows us to determine stability properties for any cubic map with real fixed points by analyzing only the Canonical Cubic Map.

Consider the cubic discrete dynamical system given in its General Form by $y_{n+1} = f_3(y_n; \lambda)$, where the iteration function is given by the following definition.

*Definition 1* (general cubic map). The *general cubic map (GCM)* is defined by

$$f_3(y) := y + P_{f_3}(y), \qquad (1)$$

where

$$P_{f_3}(y) = \alpha + \beta y + \gamma y^2 + \delta y^3 \qquad (2)$$

is called the *fixed points polynomial* of $f_3$. All the coefficients $\alpha, \beta, \gamma$, and $\delta$ are functions of the parameter $\lambda$.

It is evident that any cubic map can be put in this form by adjusting the corresponding values of the coefficients in the fixed points polynomial. By the fundamental theorem of algebra, we know that (2) has three roots, by which the GCM has three fixed points. The roots of $P_{f_3}$ are then $y_0 = S + T - (1/3)\widetilde{\gamma}$, $y_1 = -(1/2)(S+T) - (1/3)\widetilde{\gamma} + (1/2)i\sqrt{3}(S-T)$, and $y_2 = -(1/2)(S+T) - (1/3)\widetilde{\gamma} - (1/2)i\sqrt{3}(S-T)$, where $\widetilde{\alpha} = \alpha/\delta$, $\widetilde{\beta} = \beta/\delta$, $\widetilde{\gamma} = \gamma/\delta$, $Q = (3\widetilde{\beta} - \widetilde{\gamma}^2)/9$, $R = (9\widetilde{\beta}\widetilde{\gamma} - 27\widetilde{\alpha} - 2\widetilde{\gamma}^3)/54$, $D = Q^3 + R^2$, $S = \sqrt[3]{R + \sqrt{D}}$, and $T = \sqrt[3]{R - \sqrt{D}}$. The coefficients of the fixed points polynomial (2) and its roots are related by $y_0 + y_1 + y_2 = -\widetilde{\gamma}$, $y_0 y_1 + y_1 y_2 + y_2 y_0 = \widetilde{\beta}$, and $y_0 y_1 y_2 = -\widetilde{\alpha}$. $D$ is called the discriminant and we have three cases.

(i) If $D > 0$ then one fixed point is real and the other two are complex conjugates.

(ii) If $D = 0$ then the three fixed points are real with at least two of them equal.

(iii) If $D < 0$ then all fixed points are real and distinct.

It is the last two cases (real fixed points) that will interest us most for the time being. Suppose in particular that $D < 0$. Then, we can write

$$\begin{aligned}
y_0 &= 2\sqrt{-Q}\cos\left(\frac{\theta}{3}\right) - \frac{1}{3}\widetilde{\beta}, \\
y_1 &= 2\sqrt{-Q}\cos\left(\frac{\theta + \pi}{3}\right) - \frac{1}{3}\widetilde{\beta}, \qquad (3) \\
y_2 &= 2\sqrt{-Q}\cos\left(\frac{\theta + 2\pi}{3}\right) - \frac{1}{3}\widetilde{\beta},
\end{aligned}$$

where $\cos\theta = R/\sqrt{-Q^3}$. Using the previous notation we have the following definition.

*Definition 2* (linear factors form of the cubic map). Let $f_3$ be a general cubic map with three fixed points, $y_0, y_1$, and $y_2 \in \mathbb{C}$. One can write $f_3$ as

$$\begin{aligned}
h_3(y) &= y + M(y - y_0)(y - y_1)(y - y_2) \\
&= y + s\widetilde{M}(y - y_0)(y - y_1)(y - y_2),
\end{aligned} \qquad (4)$$

where all $M, y_0, y_1$, and $y_2$ are functions of the parameter $\lambda$, $s = \text{sign}(M)$, $\widetilde{M} = |M|$; one calls $h_3$ the *Linear Factors Form of the cubic map (LFFCM)*.

Now we will apply a linear transformation to (4) so that one fixed point is mapped to zero and the "amplitude" coefficient of the Linear Factors term is unity; this can be done since $f_3$ is cubic and at least one of the fixed points is real, so we can always map this fixed point to zero. The linear transformation can be chosen by each one of the following transformations:

$$T_0(x) = y_0 \pm \frac{x}{\sqrt{M}}, \qquad T_1(x) = y_1 \pm \frac{x}{\sqrt{M}},$$
$$T_2(x) = y_2 \pm \frac{x}{\sqrt{M}}, \qquad (5)$$

by taking $y = T_k(x)$, $k \in \{0, 1, 2\}$. Without loss of generality, we will use $T_0$ with the plus sign and call it simply $T$, so that we get the following.

*Definition 3* (canonical cubic map). The *Canonical Cubic Map (CCM)* is defined by

$$g_3(x; \lambda) = x + sx(x - x_1(\lambda))(x - x_2(\lambda)), \qquad (6)$$

where it has been stressed out that both fixed points $x_1$ and $x_2$ depend upon the parameter $\lambda$.

So if $M > 0$ then

$$g_3(x) = x + x(x - x_1)(x - x_2), \qquad (7)$$

and if $M < 0$

$$g_3(x) = x - x(x - x_1)(x - x_2). \qquad (8)$$

The relationship between the roots of the Linear Factors Form of the cubic map and the Canonical Cubic Map (CCM) is given by the following.

**Corollary 4.** *The fixed points of the Linear Factors Form of the cubic map and the Canonical Cubic Map are related by*

$$x_1(\lambda) = \sqrt{M(\lambda)}\left[y_1(\lambda) - y_0(\lambda)\right],$$
$$x_2(\lambda) = \sqrt{M(\lambda)}\left[y_2(\lambda) - y_0(\lambda)\right]. \quad (9)$$

We have then reduced the parametric dependence to only two functions of the parameter $\lambda$: $x_1$ and $x_2$. Notice $T$ is a homeomorphism between the *domains* of both maps; this will help us in Section 3 to prove that the Linear Factors Form and the Canonical Form of polynomial maps are actually topologically conjugate, which in turn means that the stability and chaos properties are preserved between the maps, which allows us to determine stability properties for any cubic map by analyzing only the CCM.

*2.1. Stability for the Canonical Cubic Map.* Let us determine the stability of the periodic points of the CCM. This analysis will suffice for any cubic map with real fixed points, by means of topological conjugacy. However, we can only explicitly give this for the fixed points. We already know, by construction, that the fixed points of the CCM are $x_0 = 0$, $x_1$, and $x_2$. While the first is constant, the other two fixed points are set to be functions of the parameter $\lambda$. By evaluating in $g'_3$, we get the eigenvalue functions. For $x_0 = 0$ we have $\phi_0(\lambda) = g'_3(0) = sx_1(\lambda)x_2(\lambda) + 1$. So the stability condition for this fixed point is

$$-2 < sx_1 x_2 < 0. \quad (10)$$

We can draw some conclusions from this. In order for zero to be a stable (attracting) fixed point one must have the following.

**Lemma 5.** *The following are sufficient conditions for the asymptotic stability of the zero fixed point of the Canonical Cubic Map:*

(i) *in magnitude, $|x_1||x_2| < 2$;*

(ii) *if $M > 0$, $x_1$ and $x_2$ must have different signs; or*

(iii) *if $M < 0$, $x_1$ and $x_2$ must have the same sign.*

Notice that the stability condition (10) states that the product of the relative positions from the other two fixed points to the zero fixed point must be within the range $(-2, 0)$, for positive $M$ [or $(0, 2)$ for negative $M$], for the zero fixed point to be asymptotically stable.

The case of $x_k = 0$, $k \in \{1, 2\}$, is not included in the discussion here since this would represent repeated fixed points (multiplicity), which will be discussed in Section 2.3 below; likewise, in the remainder of this section we will avoid dealing with multiplicity of the fixed points. Now, for $x_1$, its eigenvalue function is $\phi_1(\lambda) = g'_3(x_1(\lambda)) = 1 + sx_1(\lambda)(x_1(\lambda) - x_2(\lambda))$, so that the stability condition for this fixed point is

$$-2 < sx_1(x_1 - x_2) < 0. \quad (11)$$

This fact gives us the following.

**Lemma 6.** *The following are sufficient conditions for the asymptotic stability of the $x_1$ fixed point.*

*If $M > 0$ then*

(i) *$x_1$ and $x_2$ must have the same sign;*

(ii) *$|x_1| < |x_2| < |x_1 + 2/x_1|$.*

*On the other hand, if $M < 0$,*

(i) *$|x_2| < |x_1|$;*

(ii) *if $|x_1| \geq \sqrt{2}$, then $|x_1 - 2/x_1| < |x_2| < |x_1|$; or*

(iii) *if $0 < x_1 < \sqrt{2}$, then $x_1 - 2/x_1 < x_2 < x_1$; or*

(iv) *if $-\sqrt{2} < x_1 < 0$, then $x_1 < x_2 < x_1 - 2/x_1$.*

Again, notice that the stability condition (11) for $x_1$ can be translated as that the product of the relative positions between the other two fixed points and $x_1$ must be within the range $(-2, 0)$ for positive $M$ [or $(0, 2)$ for negative $M$]. Also notice that when $0 < |x_1| < \sqrt{2}$ the bound $x_1 - 2/x_1$ may be negative even if $x_1 > 0$ or positive even if $x_1 < 0$, therefore the usefulness of the distinction. For $x_2$ we have analogous results since it is indistinguishable from $x_1$ in the present formulation.

We will later generalize these "stability conditions" to functions of the parameter which are different for each fixed point, but of whose value depends on the stability of not only the fixed points, but also higher period periodic points, through period doubling bifurcations. From the stability conditions for the three fixed points we have proved the following.

**Corollary 7.** *A cubic polynomial map with three different real roots can only have a single attracting fixed point.*

*Proof.* Compare the stability conditions for the three fixed points. □

Also we have proved the following theorem.

**Theorem 8.** *Then sufficient conditions for the stability of a fixed point of the Canonical Cubic Map are as follows.*

*If $M > 0$,*

(i) *the product of the relative positions between each unstable fixed point and the stable one must be negative, which means one position is positive and the other negative, which leads us to the following;*

(ii) *the fixed point that lies between the other two will be stable, while the outer fixed points will be unstable, as long as the following holds;*

(iii) *the product of the relative positions between each unstable fixed point and the stable one must be greater than $-2$.*

*And if $M < 0$,*

(i) *the product of the relative positions between each unstable fixed point and the stable one must be positive, which means that both relative positions are positive, which leads us to the following;*

Table 1: Bifurcation values for the canonical cubic map.

| $k$ | $c_k$ |
|---|---|
| 1 | 2 |
| 2 | $3.0 \pm 0.005$ |
| 3 | $3.236 \pm 0.002$ |
| 4 | $3.288 \pm 0.0005$ |
| 5 | $3.29925 \pm 0.00025$ |
| ⋮ | ⋮ |
| $\infty$ | $\sim 3.30228 \pm 5 \times 10^{-6}$ |

(ii) *either zero or the outer fixed point will be stable, while the other two fixed points will be unstable, as long as the following holds;*

(iii) *the product of the relative positions between each unstable fixed point and the stable one must be less than 2.*

### 2.2. Higher Period Periodic Points.

Although, as previously stated, in general, we cannot calculate the values of the periodic points of period 2 or higher, we can calculate for which values of the stability conditions above the fixed points undergo period doubling bifurcations. We will see in Section 3 that these stability conditions can actually be generalized to something called the "Product Position Function," which depends on the parameter and is different for each fixed point. An asymptotic parameterization of the fixed points allowed us to determine the bifurcation values, $c_k$, of the fixed points of the CCM up to some precision. The values obtained are shown in Table 1. When the stability conditions of each fixed point cross these values, bifurcations take place. An estimation of the bifurcation value for the onset of chaos through a period doubling cascade has been calculated as $c_\infty \sim 3.30228 \pm 5 \times 10^{-6}$.

From these values, we can construct the analogue of the stability bands of the CQM for the CCM.

**Definition 9** (stability bands of the CCM). Let $x_1, x_2 : \mathcal{A} \subseteq \mathbb{R} \to \mathbb{R}$ be the two nonzero fixed points of the family of cubic maps $g_3$, as given by Definition 3, and let $\{c_k\}_{k \in \mathbb{N}}$ be the sequence of bifurcation values of Table 1. The open interval

$$(-c_{k+1}, -c_k), \quad \lambda \in \mathcal{A} \tag{12}$$

is called the *k*th *stability band* of the CCM.

Notice, however, that in contrast with the stability bands of the Canonical Quadratic Map, the stability bands of the CCM cannot be plotted along the fixed points plots, at least not directly as just defined, but rather they must be represented in a separate plot for the stability conditions, as we will see in the examples of Section 4.

### 2.3. Multiplicity of the Fixed Points.

When multiplicity of the fixed points takes place in the CCM, without loss of generality, $g_3$ can take the following forms:

$$g_3(x) = \begin{cases} x + sx^2(x - x_1), & \text{if } x_2 = x_0 = 0, x_1 \neq 0 \\ x + sx(x - x_1)^2, & \text{if } x_1 = x_2 \neq 0 \\ x + sx^3, & \text{if } x_1 = x_2 = 0, \end{cases} \tag{13}$$

with corresponding derivatives

$$g_3'(x) = \begin{cases} 1 + 2sx(x - x_1) + sx^2, & \text{if } x_2 = x_0 = 0, x_1 \neq 0 \\ 1 + sx(x - x_1)^2 \\ \quad + 2sx(x - x_1), & \text{if } x_1 = x_2 \neq 0 \\ 1 + 3sx^2, & \text{if } x_1 = x_2 = 0, \end{cases} \tag{14}$$

and therefore, $g_3'(x_k) = 1$, $k \in \{0, 1, 2\}$, for all three cases, so that we deal with nonhyperbolic fixed points.

**Proposition 10.** *The stability of the fixed points of the CCM when they present multiplicity is, without loss of generality, as follows.*

(1) *If $x_2 = x_0 = 0$, $x_1 \neq 0$, the zero fixed point is an unstable fixed point with multiplicity of two;*

  (i) *if $M > 0$ and*

    (a) *if $x_1 > 0$ it is semiasymptotically stable from the right,*
    (b) *if $x_1 < 0$ it is semiasymptotically stable from the left;*

  (ii) *or if $M < 0$ and*

    (a) *if $x_1 > 0$ it is semiasymptotically stable from the left,*
    (b) *if $x_1 < 0$ it is semiasymptotically stable from the right.*

(2) *If $x_1 = x_2 \neq 0$, this fixed point has multiplicity of two and it is unstable; moreover*

  (i) *if $M > 0$ and*

    (a) *if $x_1 > 0$ it is semiasymptotically stable from the left,*
    (b) *if $x_1 < 0$ it is semiasymptotically stable from the right;*

  (ii) *or if $M < 0$ and*

    (a) *if $x_1 > 0$ it is semiasymptotically stable from the right,*
    (b) *if $x_1 < 0$ it is semiasymptotically stable from the left.*

(3) *If $x_0 = x_1 = x_2 = 0$, the zero fixed point has multiplicity of three;*

  (i) *if $M > 0$, it is unstable;*
  (ii) *if $M < 0$, it is asymptotically stable.*

*Proof.* Notice that

$$g_3''(x) = \begin{cases} 2s(x - x_1) + 4sx, & \text{if } x_2 = x_0 = 0, x_1 \neq 0 \\ 4s(x - x_1) + 2sx, & \text{if } x_1 = x_2 \neq 0 \\ 6sx, & \text{if } x_1 = x_2 = 0, \end{cases} \quad (15)$$

$$g_3'''(x) = 6s \neq 0,$$

for all cases. Using the stability (ST) and semistability (SST) theorems for nonhyperbolic points [13, pp. 4-5], therefore

(1) if $x_2 = x_0 = 0$, $x_1 \neq 0$, the zero fixed point has multiplicity of two and we have that $g_3'(0) = 1$, $g_3''(0) = -2sx_1 \neq 0$, and $g_3'''(0) = 6s \neq 0$; therefore, by ST, the zero fixed point is an *unstable* fixed point. Applying SST we get the particular cases of semistability;

(2) if $x_1 = x_2 \neq 0$, this fixed point has multiplicity of two and we have that $g_3'(x_1) = 1$, $g_3''(x_1) = 2sx_1 \neq 0$, and $g_3'''(0) = 6s \neq 0$; therefore, by ST, this fixed point is *unstable*; moreover, the semistability cases are inferred from SST again;

(3) if $x_0 = x_1 = x_2 = 0$, the zero fixed point has multiplicity of three and we have that $g_3'(0) = 1$, $g_3''(0) = 0$, and $g_3'''(0) = 6s \neq 0$; therefore, by ST, if $M > 0$ the zero fixed point is *unstable* and if $M < 0$ it is *asymptotically stable*.

□

## 3. Polynomial Discrete Dynamical Systems

Consider again a one-dimensional discrete dynamical system defined by

$$y_{n+1} = f(y_n; \lambda), \quad (16)$$

where $f$ is a polynomial in one real variable $y$ with real fixed points and whose coefficients depend smoothly on the real parameter $\lambda$. Depending on the form of $f$ we have defined previously the General, Linear Factors, and Canonical Forms of the cubic maps. Next, we will define precisely the General and Canonical Maps of a $m$th degree polynomial map.

*Definition 11* (general polynomial map). The General Polynomial Map of $m$th degree (GPM-$m$) is defined by

$$f_m(y) := y + (-1)^{m-1} P_{f_m}(y), \quad (17)$$

where

$$P_{f_m}(y) := (-1)^{m-1} \sum_{i=0}^{m} a_i y^i. \quad (18)$$

$P_{f_m}$ is called the *fixed points polynomial* associated with $f_m$.

It is known that any $m$th degree real polynomial in one variable can be put into the General Form by means of adjusting the value of the $a_1$ coefficient properly in the fixed points polynomial. This is the broadest class of real polynomials of finite degree. The roots of $P_{f_m}$ are the fixed points of $f_m$. In the case of $m$ odd, the fundamental theorem of algebra guarantees the existence of at least one real fixed point. Let $y_i \in \mathbb{C}$, $i \in \{0, 1, \ldots, m-1\}$, be the $m$ roots of $P_{f_m}$; then $(y - y_i)$ is a factor of $P_{f_m}$ by the factor theorem; therefore we can rewrite $P_{f_m}$ as

$$P_{f_m}(y) = M(y - y_0) \cdots (y - y_{m-1})$$
$$= M \prod_{j=0}^{m-1} (y - y_j), \quad M \in \mathbb{R}, \quad (19)$$

and then define the following.

*Definition 12* (linear factors form). Let $f_m$ be the GPM-$m$ and $y_j$, $j \in \{0, \ldots, m-1\}$, its $m$ fixed points. Then one can write

$$f_m(y) = y + (-1)^{m-1} M \prod_{i=0}^{m-1} (y - y_i),$$

$$= y + (-1)^{m-1} \text{sgn}(M) |M| \prod_{i=0}^{m-1} (y - y_i), \quad (20)$$

$$= y + (-1)^{m-1} s\widetilde{M} \prod_{i=0}^{m-1} (y - y_i),$$

with the definitions of $s$ and $\widetilde{M}$ used in Section 2. One calls this the *Linear Factors Form* of $f_m$.

We can directly verify that for $m = 3$ we obtain the corresponding Linear Factors Form of the cubic maps. Once we know the $m$ fixed points of a map $f_m$, it is straightforward to write its Linear Factors Form. The motivation behind the $(-1)^{m-1}$ factor is that we want that, for purely aesthetic reasons, if $M > 0$, the fixed points are real, and $0 < y_0 < y_1 < \cdots < y_{m-1}$, we have that $f_m'(0) \geq 0$, which is thus accomplished.

We will now restrict this set of polynomials to those whose fixed points polynomials have only real roots, that is, maps with real fixed points only, though not necessarily distinct. Let us make this precise by the following.

*Definition 13* (canonical polynomials set). The Canonical Polynomials Set, denoted by $P_C[y]$, is

$$P_C[y] := \{f \in \mathbb{R}[y] \mid P_f \text{ has only real roots}\}, \quad (21)$$

where $\mathbb{R}[y]$ is the set of polynomials with real coefficients on the variable $y$. Likewise, $P_C^m[y]$ denotes $P_C[y] \cap \mathbb{R}_m[y]$, where $\mathbb{R}_m[y]$ is the set of polynomials of degree $m$ with real coefficients on the variable $y$.

The set $P_C$ has been our main work ground for the analysis in this work and, as it turns out, its elements can be put in a much nicer form, easier to understand. We can further reduce the complexity of this set of maps by means of the transformation

$$y = T_m(x) := s\widetilde{M}^{-1/(m-1)} x + y_0, \quad (22)$$

where $y_0$ is a real fixed point of the map in its linear product form. Notice that $T_m$ is linear; therefore it has an inverse

$$x = T_m^{-1}(y) = s\widetilde{M}^{1/(m-1)}(y - y_0). \quad (23)$$

$T_m$ is in fact a homeomorphism. We will drop the subscript $m$ when referring to the transformation for no specific degree. Applying this transformation to $y$ we reach the following.

*Definition 14* (canonical polynomial map). The canonical polynomial map of $m$th degree (CPM-$m$) is

$$g_m(x) := x + (-1)^{m-1} s^m x \prod_{i=1}^{m-1}(x - x_i), \quad m \geq 2, \quad (24)$$

where

$$x_i = s\widetilde{M}^{1/(m-1)}(y_i - y_0), \quad (25)$$

and $y_j$ are the $m$ fixed points of the corresponding Linear Factors Form map of $m$th degree, (at least) $y_0$ being real.

It is clear from the definition that $x_0 = 0$ always. Notice also that the $x_i$ result from evaluating $T_m^{-1}$ in the corresponding $y_i$. We can easily prove that not only does the canonical map result from applying $T$, but also the canonical map is in fact $T$-conjugate to the Linear Factors Form.

**Proposition 15.** *Let $f_m$ and $T_m$ and $g_m$ be as defined above, having $f_m$ at least one real fixed point; let $y_0$ be this real fixed point, without loss of generality. Then $f_m$ is $T_m$-conjugate to $g_m$.*

*Proof.* It is clear that $T_m$ is a homeomorphism since it is linear. Then, we must only prove that $T_m \circ f_m = g_m \circ T_m$; that is, $f_m(T_m(x)) = T_m(g_m(x))$. We then have

$$f_m(T_m(x))$$
$$= f_m\left(s\widetilde{M}^{-1/(m-1)}x + y_0\right)$$
$$= s\widetilde{M}^{-1/(m-1)}x + y_0$$
$$\quad + (-1)^{m-1} s\widetilde{M} \prod_{i=0}^{m-1}\left(s\widetilde{M}^{-1/(m-1)}x + y_0 - y_i\right)$$
$$= s\widetilde{M}^{-1/(m-1)}x + (-1)^{m-1} s^2 \widetilde{M}\widetilde{M}^{-1/(m-1)}x$$
$$\quad \cdot \prod_{i=1}^{m-1}\left(s\widetilde{M}^{-1/(m-1)}x + y_0 - y_i\right) + y_0$$
$$= s\widetilde{M}^{-1/(m-1)}x + (-1)^{m-1} s^{m-1} \widetilde{M}^{-1/(m-1)}x$$
$$\quad \cdot \prod_{i=1}^{m-1}\left[x - s\widetilde{M}^{1/(m-1)}(y_i - y_0)\right] + y_0$$

$$= s\widetilde{M}^{-1/(m-1)}x + (-1)^{m-1} s^{m-1} M^{-1/(m-1)} x$$
$$\quad \cdot \prod_{i=1}^{m-1}(x - x_i) + y_0$$
$$= s\widetilde{M}^{-1/(m-1)}\left[x + (-1)^{m-1} s^m x \prod_{i=1}^{m-1}(x - x_i)\right] + y_0$$
$$= T_m(g_m(x)), \quad (26)$$

where we have used $s^2 = 1$ and $s = s^{-1}$. □

This turns out to be very useful, since we know that topological conjugacy is an equivalence relation that *preserves the property of chaos*. This means that the analysis of stability and chaos (i.e., the "dynamics") of real polynomial maps with real fixed points is reduced to the study of the canonical polynomial maps defined above, since we can always take any polynomial in $P_C[x]$ to its Canonical Form by means of $T$, determine the stability properties, and then go back to the original polynomial. A commutative diagram of the conjugacy is in Figure 5.

*3.1. Stability and Chaos in the Canonical Map of Degree m.* The derivative of $g_m$, recalling $x_0 = 0$ to simplify notation, is

$$g_m'(x) = 1 + (-1)^{m-1} s^m \sum_{j=0}^{m-1} \prod_{i=0, i \neq j}^{m-1}(x - x_i). \quad (27)$$

Evaluating (27) in the fixed point $x_k$ we get the eigenvalue function for each $x_k$:

$$\phi_k(\lambda) = g_m'(x_k(\lambda))$$
$$= 1 + (-1)^{m-1} s^m \prod_{i=0, i \neq k}^{m-1}(x_k(\lambda) - x_i(\lambda)) \quad (28)$$
$$= 1 + s^m \prod_{i=0, i \neq k}^{m-1}(x_i(\lambda) - x_k(\lambda)).$$

Then, the asymptotic stability condition $|g_m'(x_k)| < 1$ implies that

$$-2 < s^m \prod_{i=0, i \neq k}^{m-1}(x_i - x_k) < 0. \quad (29)$$

From (29) we can recover all the stability conditions for the fixed points of the Canonical Quadratic Map and cubic map. The above leads us to the following.

*Definition 16* (Product Position Function). Let $g_m$ be the canonical polynomial map of $m$th degree and $x_0 = 0$, and let $x_1, \ldots, x_{m-1}$ be its $m$ fixed points, all of which depend upon the parameter $\lambda$. Let $x_k$ be a real fixed point among the latter. Then

$$D_{m,k}(\lambda) := s^m \prod_{i=0, i \neq k}^{m-1} (x_i(\lambda) - x_k(\lambda)), \quad (30)$$

$$k \in \{0, \ldots, m-1\}, \quad m \geq 2,$$

is called the *Product Position Function (PPF)* of $x_k$.

The definition is motivated by the fact that $D_{m,k}$ is a product of the positions relative to $x_k$ of each of the other $m - 1$ fixed points and that this quantity is fundamental in determining the stability of the fixed points. These positions are positive when $x_i > x_k$ and negative when $x_i < x_k$. We have stressed the dependence on the parameter $\lambda$ in the definition of $D_{m,k}$ so that its character as a function is clear, stemming from the corresponding dependence on $\lambda$ of the fixed points. In this way, the stability condition for the $k$th fixed point is reduced to

$$-2 < D_{m,k}(\lambda) < 0. \quad (31)$$

Since $D_{m,k}$ must be negative in order for $x_k$ to be stable as a sufficient condition and an *odd* number of factors $(x_i - x_k)$ must be negative for the product in $D_{m,k}$ to be negative, it follows that if $m$ is even or $M > 0$, an odd number of negative factors $(x_i - x_k)$ is a necessary condition for the hyperbolic fixed point $x_k$ to be stable; that is, if $M > 0$, an *odd* number of fixed points must lie below $x_k$ and, consequently, an even or zero (resp., odd) number of fixed points must lie above $x_k$ if $m$ is even (resp., odd). By similar arguments, we can prove the following.

**Proposition 17** (necessary conditions for the stability of $x_k$). *Let $g_m$, $D_{m,k}$ be defined as above and let $x_k$ be a hyperbolic real fixed point of $g_m$. The following are necessary conditions for $x_k$ to be an asymptotically stable fixed point:*

(i) *if $M > 0$ or $m$ is even, an odd number of fixed points must have values lower than $x_k$; or*

(ii) *if $m$ is odd and $M < 0$, zero or an even number of fixed points must have values lower than $x_k$.*

We must remark that the above conditions are *not* sufficient for a fixed point to be an attractor. The sufficient condition, however, is stated as follows.

**Theorem 18** (sufficient condition for the stability of $x_k$). *Let $g_m$, $D_{m,k}$ be defined as above and let $x_k$ be a hyperbolic real fixed point of $g_m$. Then, a necessary and sufficient condition for $x_k$ to be an attractor is that*

$$-2 < D_{m,k}(\lambda) < 0. \quad (32)$$

Below the value of $-2$ there are other "stability bands" that lead to further period doubling bifurcations of the fixed points as they are crossed, but they must be calculated numerically and, as we have seen, depend on the degree $m$ of the polynomial.

Let us remark that for those values of $\lambda$ in (32) where $D_{m,k}(\lambda) = -2$ or $D_{m,k}(\lambda) = 0$ there are some stability conditions that require higher parametric derivatives; see, for instance, [14, 15].

## 4. Examples

Here we will deal with specific parameterizations for the fixed points $x_1$ and $x_2$ in order to clarify the above findings and to demonstrate how we can construct bifurcation diagrams with specific predetermined properties with cubic maps. We will consider $M > 0$ unless otherwise stated explicitly.

*Example 1.* First, consider linear parameterizations for both $x_1$ and $x_2$ as

$$x_1(\lambda) = -\lambda, \quad x_2(\lambda) = \lambda. \quad (33)$$

The result is plotted in the lower panel of Figure 1, where we see that the middle fixed point is $x_0 = 0$ always, so we expect this to be the only stable fixed point, until the separation between this and the other points breaks the stability condition and the period doubling bifurcation cascade sets on. The corresponding stability conditions are shown in the middle panel of Figure 1, where we confirm that the curve for $x_0$ is the only one within the stability band $(-2, 0)$ until $\lambda \approx 1.45$, where the curve crosses the barrier of $-2$ getting into the stability band $(-3, -2)$, causing $x_0$ to bifurcate. The corresponding bifurcation diagram is shown in the upper panel of Figure 1, where we confirm the statement stated above.

*Example 2.* Now we will explore the full range of stability regions by making a linearly varying fixed point pass through the regions defined by the constant $x_0$ and a constant $x_1$. We define then

$$x_1(\lambda) = 2, \quad x_2(\lambda) = 6\lambda + 1. \quad (34)$$

We then obtain the plot of the lower panel of Figure 2, for the selected range of interest of the parameter $\lambda$. In the middle panel of the same figure we can see the stability curves for the fixed points, where we see that initially, from left to right, all fixed points are unstable and then, progressively, $x_0$, $x_2$, and $x_1$ become stable, the latter one losing stability for still greater values of $\lambda$. The corresponding bifurcation diagram is shown in the upper panel, where we can see how first the stable fixed point is $x_0 = 0$, since it is the middle one, but begins in the chaotic region and goes "reversal" towards being stable; then, as $x_2$ crosses through zero, it becomes the middle stable fixed point and when it in turn crosses the constant $x_2$, this latter one becomes the stable fixed point, again loosing stability when $x_2$ crosses the stability band for $x_1$.

*Example 3* (quartic maps). Using Definition 12 for $n = 4$, we have that $f_4(y) = y - M(y - y_0)(y - y_1)(y - y_2)(y - y_3)$. Suppose $f_4$ has at least one real fixed point. Without loss of generality, suppose this fixed point is $y_0$. Then $T_4(x) = s\widetilde{M}^{-1/3}x + y_0$. Making the substitution $y = T_4(x)$ we can verify that we get $g_4(x) = x - x(x - x_1)(x - x_2)(x - x_3)$, where $x_i = s\widetilde{M}^{1/3}(y_i - y_0)$, $i \in \{0, 1, 2, 3\}$. The stability of a real fixed point $x_k$ is given by the Product Position Function $D_{4,k}(\lambda) = \prod_{i=0, i \neq k}^{3}(x_i(\lambda) - x_k(\lambda))$, whose value must remain between minus two and zero in order for $x_k$ to be asymptotically stable; that is, if all fixed points are real,

$$\begin{aligned}
-2 &< D_{4,0}(\lambda) = x_1 x_2 x_3 < 0, \\
-2 &< D_{4,1}(\lambda) = -x_1(x_2 - x_1)(x_3 - x_1) < 0, \\
-2 &< D_{4,2}(\lambda) = -x_2(x_1 - x_2)(x_3 - x_2) < 0, \\
-2 &< D_{4,3}(\lambda) = -x_3(x_1 - x_3)(x_2 - x_3) < 0,
\end{aligned} \quad (35)$$

for 0, $x_1$, $x_2$, and $x_3$ to be asymptotically stable fixed points, respectively. For example, let

$$x_1(\lambda) = \lambda, \quad x_2(\lambda) = -\lambda, \quad x_3(\lambda) = 2\lambda. \quad (36)$$

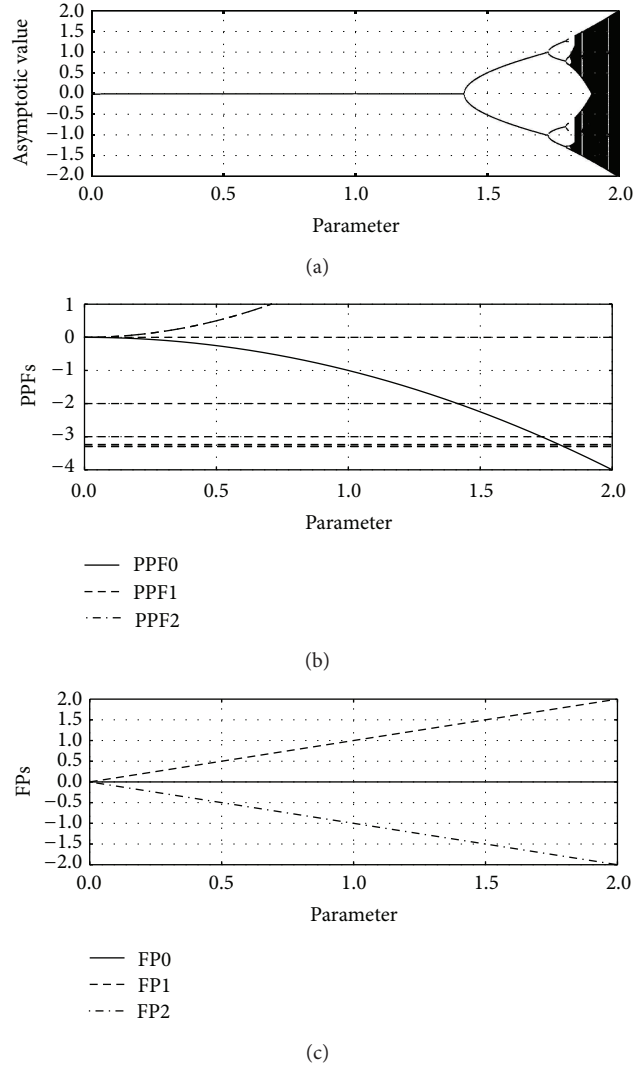

Figure 1: Bifurcation diagram (a), Product Position Functions and stability bands (b), and fixed points (c) for the linear parameterization of the fixed points example of the CCM.

The plots of these fixed points with their corresponding parametric dependence on $\lambda$ are shown in Figure 3(c). In light of Proposition 17 we expect only $x_0$ and $x_3$ to be able to be asymptotically stable fixed points in any given range of $\lambda$. As Figure 3(b) shows, precisely $x_0$ and $x_3$ are the fixed points whose product distances cross the stability band $(-2, 0)$ in the range of $\lambda$ being plotted. As we recall, the product distance functions are the "stability conditions" of the fixed points. As long as the product distances remain within the stability interval, the fixed points are attractors, as we can verify in Figure 3(a); also in this last panel, we can see the two attracting fixed points at the beginning of the plotted range; then, first $x_3$ loses its stability and gives rise to the period doubling bifurcations cascade which leads to chaotic behavior; later, zero also loses its stability and also gives rise to period doubling and chaos.

*Example 4.* The logistic map, $L_\lambda(x) = \lambda x(1 - x)$, is the most immediate and obvious example application [16–19]. This map undergoes a series of period doubling bifurcations starting at the value of $\lambda = 3$, ultimately achieving a chaotic nature at $\lambda \approx 3.570$ [18, p. 47]. The fixed points of the logistic map are $y_0 = 0$ and $y_1 = (\lambda - 1)/\lambda$ [18, p. 43]. The corresponding Linear Factors Form of the logistic map is then $h_\lambda(x) = x - \lambda x(x - ((\lambda - 1)/\lambda))$, where we can identify the functions of the parameters $M$, $y_0$, and $y_1$ from its definitions [13] as $s = +1$, $\widetilde{M}(\lambda) = \lambda$, $y_0(\lambda) = 0$, and $y_1(\lambda) = (\lambda - 1)/\lambda$. The corresponding nonzero fixed point of the canonical logistic map, $x_1(\lambda) = s\widetilde{M}(y_1 - y_0)$, is then simply [13] $x_1(\lambda) = \lambda - 1$, from which we can state that the *canonical logistic map* takes the explicit form $g_\lambda = x - x(x - \lambda + 1)$. In order to determine the stability properties of these fixed points, both zero and nonzero, in the canonical logistic map, it is then sufficient, as we have proved in Section 3, to observe the behavior of the Product Position Functions (PPFs) of these fixed points; namely, $D_{g,0}(\lambda) = x_1 - 0 = x_1 = \lambda - 1$ and $D_{g,1}(\lambda) = 0 - x_1 = -x_1 = 1 - \lambda$. By determining when these PPFs cross the stability bands whose boundaries are shown in Table 2 [13] we can readily determine when these fixed points are stable or unstable, when they bifurcate, and when they reach any $2^n$ attracting periodic orbit for any $n$, up to crossing the $b_\infty$ band. This whole process is depicted in Figure 4. In particular, we can see from Table 2, and again in Figure 4, that when $-1 < x_1 < 0$, the zero fixed point is attracting since its PPF lies within the first stability band and then exchanges stability at $x_1 = 0$, when this last FP becomes stable and proceeds to a period doubling cascade upon its PPF, $-x_1$, crossing the bands defined by the bifurcation values $-b_1$, $-b_2$, and so forth until reaching $x_1 = b_\infty \approx 2.569941$. This last value agrees quite well and improves upon the approximation reported in [18] of 3.570 for the logistic map, since with the calculations of the present work $\lambda_\infty = 1 + b_\infty \approx 3.569941 \pm 5 \times 10^{-7}$. Finally, since these maps are topologically conjugate and it is known that the logistic map is chaotic starting with $\lambda = 4$ [18], we conclude that the Canonical Quadratic Map must be so starting from $x_1 = 3$, which we may denote by $b_c$, our final "bifurcation" value.

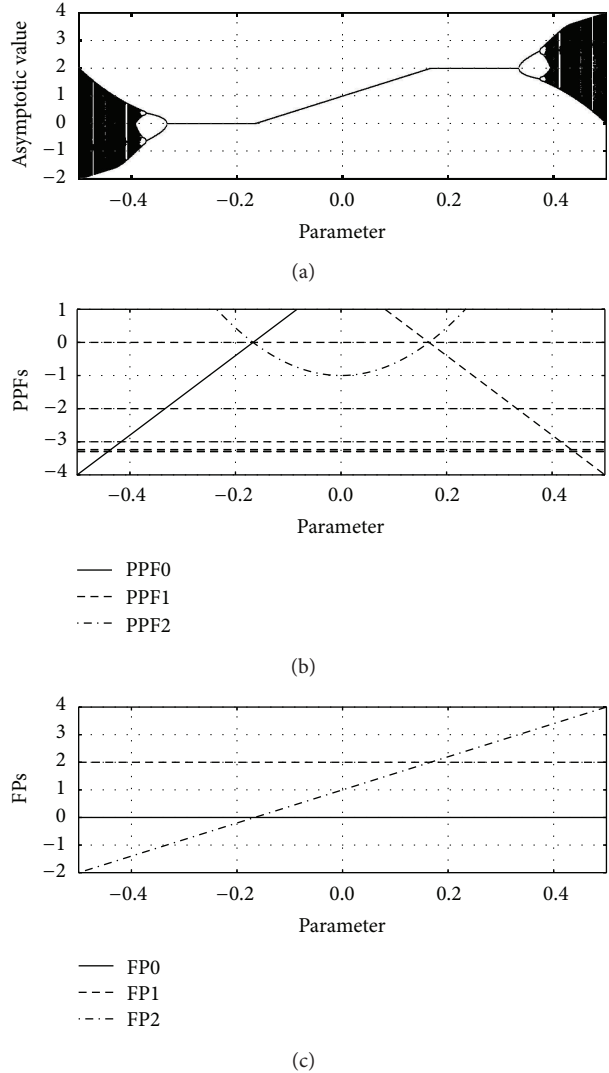

Figure 2: Bifurcation diagram (a), Product Position Functions and stability bands (b), and fixed points (c) for the constant and linear mixed parameterizations of the fixed points example of the CCM.

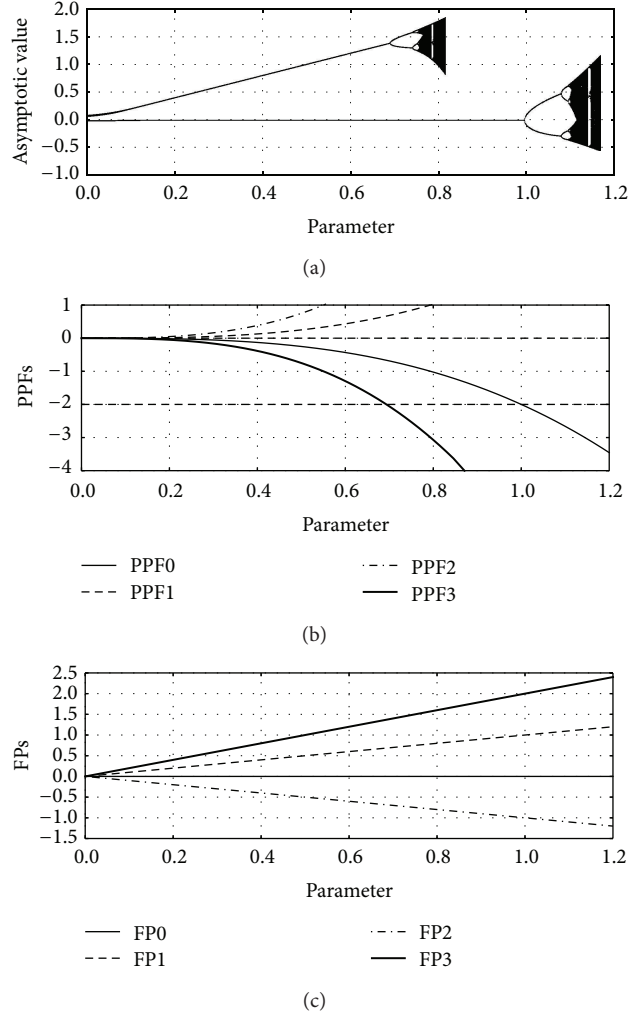

Figure 3: Bifurcation diagram (a), Product Position Functions and first stability band (b), and fixed points (c) of the quartic polynomial map example.

Table 2: Bifurcation values for the CQM. Reproduced from [13].

| $k$ | $b_k$ |
| --- | --- |
| 1 | 2 |
| 2 | $\sqrt{6}$ |
| 3 | $2.5440 \pm 0.0005$ |
| 4 | $2.5642 \pm 0.0002$ |
| 5 | $2.56871 \pm 4 \times 10^{-5}$ |
| 6 | $2.56966 \pm 1 \times 10^{-5}$ |
| 7 | $2.569881 \pm 5 \times 10^{-6}$ |
| $\vdots$ | $\vdots$ |
| $\infty$ | $\sim 2.569941 \pm 5 \times 10^{-7}$ |

*Example 5* (harvesting strategies). The connection of the logistic map with population models is old and well known. In [19] there are a few examples of second degree polynomials used as recurrence functions for modeling "harvesting," or hunting, strategies of animal populations. The main idea is that the animal populations grow whenever there are food and resources in the environment which, by account of its finite resources, has a certain "carrying capacity"; this leads to a maximum population which this environment can hold, the population growing according to the logistic model. The population may then be "harvested," or hunted, at a certain rate yet to be specified and, depending on this rate, it is not hard to imagine that the final fate of the population of animals may be (i) extinction, if the rate is too high; (ii) steady population below the carrying capacity, if the rate is "just right"; or (iii) steady population at its maximum value dictated by the carrying capacity of the environment. A final fourth possibility—perhaps more rare—is to (iv) bring more individuals of the species from outside the system under analysis and introduce them to it, therefore making it possible for the population to surpass in number the carrying capacity of the system, but only to return naturally to the maximum value after a finite number "time-steps." To examine this in detail consider the system defined by the recurrence relation $\Delta y \equiv y_{n+1} - y_n = r(1 - y_n)y_n$. Here the population growth in any

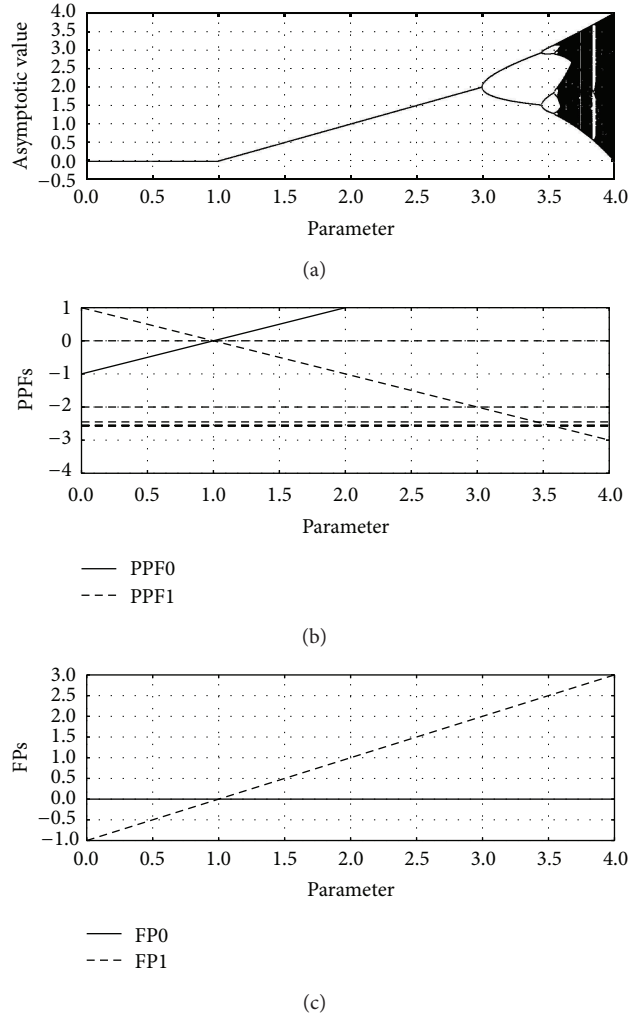

FIGURE 4: Bifurcation diagram (a), Product Position Functions (b), and fixed points (c) for the canonical logistic map.

$$\begin{array}{ccccc}
y & = & y & \xrightarrow{f_m} & f_m(y) & = & f_m(y) \\
T \uparrow & & \downarrow T^{-1} & & T \uparrow & & \downarrow T^{-1} \\
x & = & x & \xrightarrow{g_m} & g_m(x) & = & g_m(x)
\end{array}$$

FIGURE 5

given period $\Delta y$ is proportional to both the initial population $y_n$ and the difference between this and the maximum population, normalized to the value of 1. The proportion constant is the growth rate $r$. After some simplification, we can rewrite system as $y_{n+1} = (1 + r)y_n - ry_n^2$. We have yet to add the "harvesting term(s)," which we might do in several ways. If we consider a fixed rate (say each period we harvest a proportion $b$ of the population (in terms of the maximum)), then $f(y) = (1 + r)y - ry^2 - b$ is the corresponding map of this discrete dynamical system. From here we see that the fixed points polynomial (FPP) of $f$ is $P_f(y) = ry^2 - ry + b$, whose FPs we can determine to be $y_\pm = (1 \pm \sqrt{1 - 4b/r})/2$, from which we immediately calculate the Linear Factors Form (LFF) to be $f(y) = y - ry(y - y_+)(y - y_-)$. From this form it is also straightforward to determine that in the corresponding Canonical Form we have $x_1(b) = r(y_+ - y_-) = \sqrt{r(r - 4b)}$. Both systems are related then by the linear transformations $y_n = y_- + x_n/r$ and $x_n = r(y_n - y_-)$. With this choice, the zero fixed point in the canonical map corresponds to $y_-$ and the nonzero fixed point to $y_+$. To analyze the stability of this system, consider $r$ to be fixed and given and consider $b$ to be the parameter of this family of systems. The one immediate conclusion is that, for $x_1 \in \mathbb{R}$, we necessarily have that $x_1 \geq 0$. Now, real $x_1$ implies $b \leq r/4$. Over the $r/4$ value $x_1$ becomes complex which does not give any fixed points (but would mean "overharvesting"). Since "harvesting" cannot be negative, it is clear $b = 0$ corresponds to the maximum value of $x_1 = r$ and that $x_1 = 0$ when $b = r/4$. Remember now that to analyze the stability of $x_1$ we consider its PPF; $D_1 = -x_1$. Analogously, $D_0 = x_1$. Since the maximum value of $x_1$ is $r$, this being the "unconstrained growth rate," $0 < r < 1$, we have that $-r < D_1 < 0$, therefore putting the $x_1$ in the stability range between $b_0 = 0$ and $-b_1 = -2$ determined by the first stability band. This range is never left in any situation with physical meaning so we conclude that the nonzero fixed point, that is, $y_+$ in the original system, is always the only asymptotically stable fixed point and the zero fixed point, that is, $y_-$ in the original system, is always unstable. The only case to analyze with care is $b = 0$ since then the two fixed points collide, but the semistability theorem [18] guarantees that in this case $x_1 = 0$ is semistable from the right.

In conclusion,

(1) when $b = r/4$ the population faces extinction asymptotically. Over this value extinction is achieved in a finite number of steps, there not being any more fixed points;

(2) when $0 < b < r/4$, $x_n \to r$ and the population tends to $y_+ = 0.5(1 + \sqrt{1 - 4b/r})$;

(3) when $b = 0$, that is, no harvesting, $x_n$ still tends to $x_1 = 0$ from the right and, correspondingly, the population tends asymptotically to $y_+ = 1$.

## 5. Conclusions

We can summarize the findings of this work as having successfully given conditions for the stability of the fixed points of any real polynomial map with real fixed points and that depends on a single parameter. In order to do this we have defined "canonical polynomial maps" which are topologically conjugate to any polynomial map of the same degree with real fixed points. Then, the stability of the fixed points of the canonical polynomial maps has been found to depend solely on a special function called "product position" of a given fixed point. The values of this product position determine the stability of the fixed point and when it bifurcates to give rise to attracting periodic orbits of period $2^n$ for all $n$ and even when chaos arises through the period doubling cascade, as it passes through different "stability bands," although the exact values and widths of these stability bands are yet to be calculated for regions of type greater than one for higher order polynomials. The latter must be done numerically. The proposed methodology allows us to create discrete dynamical systems with some prescribed bifurcation diagram. Ultimately it is desired to obtain extensive tables of the bifurcation values for higher order polynomials. The power and simplicity of the proposed methodology will best be appreciated with 3rd or higher degree polynomials and when the implications for the Taylor polynomial of any nonlinear map are understood.

## Conflict of Interests

The authors declare that there is no conflict of interests regarding the publication of this paper.

## Acknowledgments

The authors wish to thank the support of the Mexican National Council of Science and Technology (CONACYT) and the Centro de Investigación en Matemáticas (CIMAT) for their financial support for the present research.

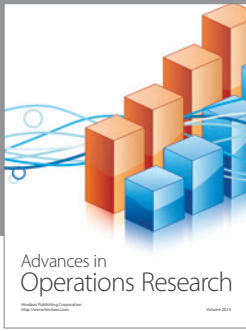 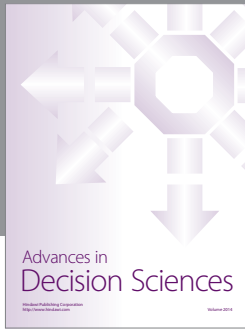 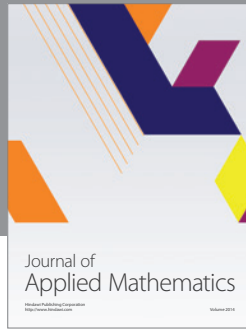 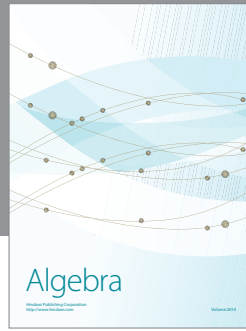 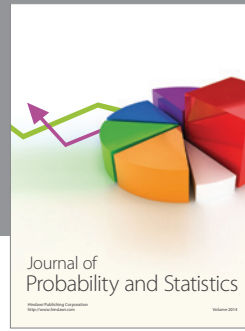
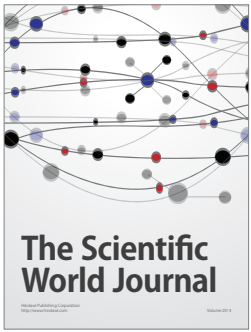 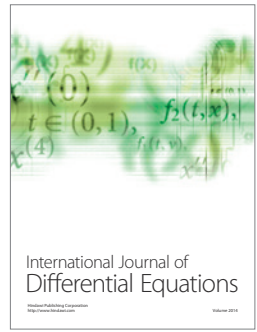
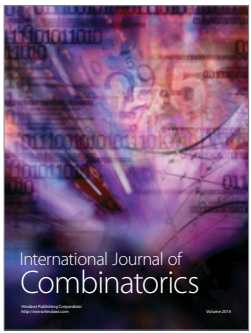 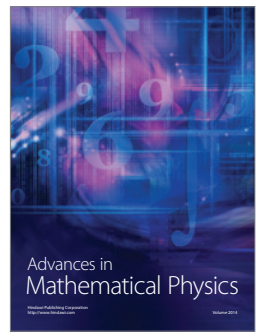
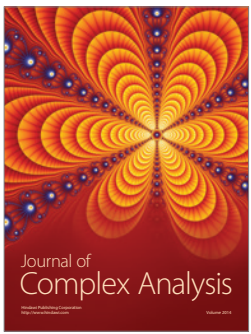 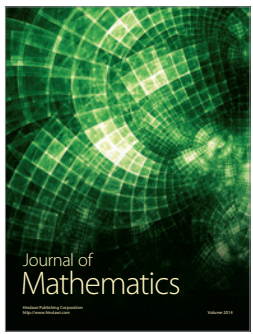 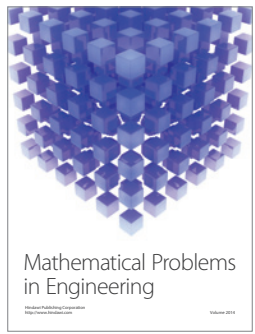 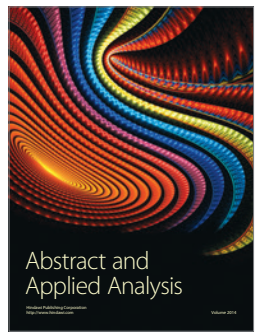 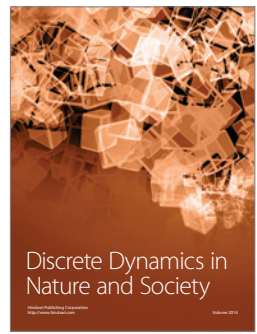
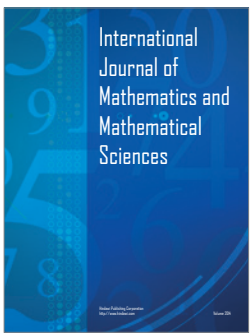 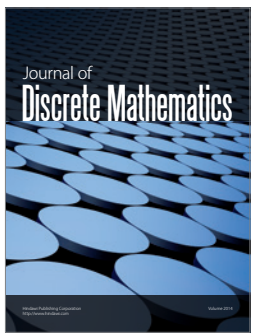 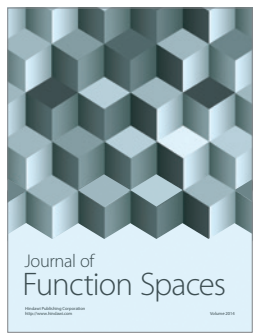 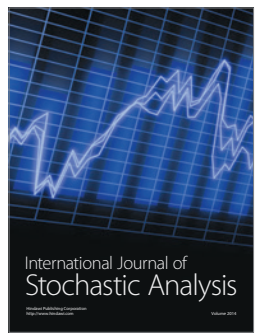 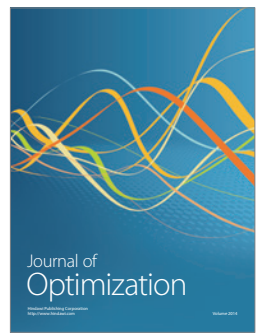